\documentclass[sigconf,authorversion,screen,nonacm]{acmart} %For preprint

\usepackage{pdfpages}
\usepackage{enumitem}

\usepackage{graphicx}

\usepackage{eso-pic}
\usepackage{xcolor}
\usepackage{ifthen}

\newboolean{showheader}
\setboolean{showheader}{true} % Set to false to not show the header
\newcommand\PlaceHeader{%
  \AtPageUpperLeft{%
    \put(\LenToUnit{0.09\paperwidth},\LenToUnit{-1cm}){%
      \parbox{1.0\textwidth}{%
        \centering\textbf{\textcolor{blue}{\href{https://conf.researchr.org/home/icse-2024/wsese-2024}{Accepted at the International Workshop on Methodological Issues in Empirical Software Engineering (WSESE2024), ICSE 2024.}}}%
      }%
    }%
  }%
}
\AddToShipoutPictureBG{%
  \ifthenelse{\boolean{showheader}}{\PlaceHeader}{}
}
\copyrightyear{2024}
\acmYear{2024}
\setcopyright{rightsretained}
\acmConference[WSESE '24 ]{International Workshop on Methodological Issues with Empirical Studies in Software Engineering}{April 16, 2024}{Lisbon, Portugal}
\acmBooktitle{International Workshop on Methodological Issues with Empirical Studies in Software Engineering (WSESE '24 ), April 16, 2024, Lisbon, Portugal}\acmDOI{10.1145/3643664.3648208}
\acmISBN{979-8-4007-0567-0/24/04}

\begin{document}

\title{Insights Towards Better Case Study Reporting in Software Engineering}

\author{Sergio Rico}
\orcid{0000-0002-9348-2912}
\affiliation{%
\department{Department of Communication, Quality Management, and Information Systems }
  \institution{Mid Sweden University}
  \city{Östersund}
  \country{Sweden}
}
\email{sergio.rico@miun.se}

\begin{abstract}

  Case studies are a popular and noteworthy type of research study in software engineering, offering significant potential to impact industry practices by investigating phenomena in their natural contexts. 
  This potential to reach a broad audience beyond the academic community is often undermined by deficiencies in reporting, particularly in the context description, study classification, generalizability, and the handling of validity threats. 
  This paper presents a reflective analysis aiming to share insights that can enhance the quality and impact of case study reporting.
    
  We emphasize the need to follow established guidelines, accurate classification, and detailed context descriptions in case studies.
  Additionally, particular focus is placed on articulating generalizable findings and thoroughly discussing generalizability threats. 
  We aim to encourage researchers to adopt more rigorous and communicative strategies, ensuring that case studies are methodologically sound, resonate with, and apply to software engineering practitioners and the broader academic community. The reflections and recommendations offered in this paper aim to ensure that insights from case studies are transparent, understandable, and tailored to meet the needs of both academic researchers and industry practitioners. In doing so, we seek to enhance the real-world applicability of academic research, bridging the gap between theoretical research and practical implementation in industry.

\end{abstract}


\maketitle

\section{Introduction}
Case studies have been recognized as one of the three well-established empirical methods in software engineering, particularly suited for investigating phenomena in their natural, real-world contexts~\cite{felderer2020evolution}. 
Google Scholar returns more than 5000 papers citing the guidelines for case study research by Runeson and Höst~\cite{runeson2009guidelines}.
Case studies, offering insights and evaluations from real-world scenarios, have a considerable potential to influence and guide practitioners.
However, a critical challenge arises from the reader's perspective: discerning the specific context of the study and determining the extent to which its findings apply to other scenarios. 
While case studies are not designed for statistical generalization, they can be analytically generalizable~\cite{runeson2009guidelines}.
The findings, context, and identification of validity threats are critical for determining the applicability of case study results to other contexts. Meticulous reporting in articulating these elements may enhance their clarity in academic discourse and utility in practical scenarios.

As a relatively young discipline, %recently celebrating its 50th anniversary, 
software engineering is continuously evolving in its methodological approaches~\cite{felderer2020evolution}. This evolution underscores the growing recognition of the need for open discussions and reflective analysis on methodological issues, deemed essential for the field's advancement~\cite{ralph2020empirical}. However, the concerns about effectively communicating empirical research findings in software engineering, as highlighted by Shaw~\cite{shaw2003writing} and Feldt et al.~\cite{feldt2010validity}, persist as an ongoing challenge.
Despite these calls to action, effectively addressing these concerns remains a pressing issue. This ongoing challenge serves as a reminder of the fundamental commitment of software engineering research — to deliver tangible value to software practice. Such a situation draws attention to the critical need for ongoing refinement of research methods. This refinement is not just about advancing academic knowledge but is also crucial for ensuring a significant practical impact.

This paper contributes to the ongoing dialogue in software engineering research, focusing on enhancing case study reporting.
By revisiting the \textbf{key aspects: guidelines and classification, context, findings, and generalizability threats}, we aim to highlight areas often overlooked in case study reports. 
We aim to offer constructive insights and suggestions to bridge the gap between academic research and practical application. 
In doing so, we aspire to ensure that the insights derived from case studies are academically rigorous and broadly applicable, enhancing their relevance and impact in practice. The takeaways (Table \ref{tab:analysis_advice}.) call researchers to critically reassess and refine their reporting strategies, elevating the significance and applicability of case studies.

\section{Key Aspects Explored}
\label{sec:key_topics}

The elements guiding our reflections are rooted in the existing literature and driven by our conviction that software engineering research can and should be more relevant and applicable to industry practices.
%We aim to bridge the often observed gap between theoretical research and its real-world implementation, ensuring that the insights and contributions of software engineering research are effectively communicated and leveraged in industry settings.

\subsection{Guidelines and Classification as Case Studies}
\label{sec:classification}

In software engineering, there are established guidelines for conducting and reporting case studies. Runeson and Höst seminal work provides comprehensive guidelines that have become a key reference in case study methodology~\cite{runeson2009guidelines}. These guidelines synthesize an extended book that provides a detailed exploration of case study research~\cite{runeson2012case}. Additionally, there are ongoing efforts within the research community to develop common standards for empirical research, including case studies~\cite{ralph2021acm}. These guidelines ensure that case studies are well-structured, clear, and focused on exploring phenomena in real-world contexts.

Wohlin and Rainer's critical analysis of case study usage in software engineering research reveals a significant inconsistency in how the term ``case study'' is employed~\cite{wohlin2022case}. Their study scrutinizes articles citing case study guidelines and reports that only about 50\% of the studies labeled as ``case study'' are correctly identified. Intriguingly, they find that approximately 40\% of these supposed case studies are more accordingly categorized as ``small-scale evaluations''. To address this, Wohlin and Rainer developed a checklist, a self-assessment scheme, and a smell indicator to ensure correct labeling and encourage accurate use of the term ``case study''.

The insights derived from Wohlin and Rainer’s work are highly pertinent to the objectives of our reflection. 
We advocate for their call for increased rigor and awareness in reporting and classification for two reasons.
Firstly, accurately classifying studies as case studies enhances their potential for transferability of findings to various contexts, particularly in industry settings, where practitioners can get valuable insights from well-documented case studies. Secondly, a more accurate classification of case studies aids in conducting secondary studies, such as systematic literature reviews, case study surveys, and meta-analyses, by providing a more precise and reliable foundation for primary studies. A more disciplined approach to classifying case studies in our field could thus substantially improve the efficacy of communication, whether in peer review processes or the synthesis of research through secondary studies.

\subsection{Context Reporting}
\label{sec:context}

Understanding the context in which a case study is conducted is essential for researchers and practitioners to evaluate the evidence and its applicability to their contexts. 
Briand et al.'s work emphasizes a shift to context-driven research in software engineering, advocating for addressing specific, real-world problems in collaboration with industry partners~\cite{briand2017case}.
Dybå et al. further highlight the central role of context in empirical software engineering, critically analyzing how overlooking context can impact the validity and applicability of research findings~\cite{dybaa2012works}. 
Petersen et al. contribute by focusing on developing a structured approach for context description, crucial for evidence aggregation and decision-making regarding the applicability of solutions in software engineering~\cite{petersen2009context}.
These papers underscore the need for contextual understanding in research, setting the stage for our analysis of case study reporting.

We consider the reporting of context as a critical aspect of analysis. The depth and clarity with which context is reported in case studies are essential for several reasons.
It allows researchers and practitioners to understand how findings might be applicable in similar contexts.
A well-documented context also facilitates evidence aggregation, enhancing the robustness and reliability of systematic literature reviews and other secondary studies. 
Furthermore, a comprehensive context description is crucial for appraising evidence and comparing studies, providing practitioners and researchers with the necessary information to assess the applicability of findings to their specific contexts.
This emphasis on context aligns with our paper's objective to discuss the interpretability and applicability of case study research in software engineering, ensuring that it maintains academic rigor and holds practical significance.

\subsection{Generalizable Findings and Threats to External Validity}
\label{sec:validity}

Understanding generalizable findings in case studies is crucial to enhancing their value and applicability. These findings, essentially the learnings or contributions of a case study, are what can be potentially applied to other contexts. 
While case studies in software engineering do not aim for statistical generalizability, they can be analytically generalizable~\cite{runeson2009guidelines}.
As Runeson et al.~\cite{runeson2009guidelines} discussed, this concept refers to applying findings to cases with common characteristics.
An essential step in achieving this is the explicit report of what the authors consider their study's key learnings or generalizable findings. Such clarity enhances the comprehensibility of the case study and aids in determining its broader relevance and applicability. 

In line with this, identifying and discussing threats to external validity is critical for ensuring the applicability of case study findings.
Runeson et al. emphasize the importance of evaluating how far case study findings can be extended to other contexts, focusing on the relevance of these findings beyond the specific case studied~\cite{runeson2009guidelines}.
This evaluation is crucial for enabling analytical generalization in case studies, aiming to apply results to cases with common characteristics. 
A recently published reflection by Verdecchia et al.~\cite{verdecchia2023threats} on threats to validity in software engineering research underscores recurrent shortcomings, particularly in treating generalizability issues. 
Their work contributes to a meaningful discussion on addressing these threats, advocating for a more systematic approach to documenting and evaluating generalizability in empirical studies.

Examining generalizability threats takes center stage, considering its critical role in determining the broader applicability of case study findings. 
The reflection by Verdecchia et al.~\cite{verdecchia2023threats} on the often superficial treatment of such threats in software engineering research aligns with our focus on enhancing the reporting quality of case studies. 
Their observations stress the need for a context-rich and detailed discussion of threats to generalizability, essential for ensuring that the findings of a case study are not only academically sound but also practically applicable and relevant in varied settings.
This approach is critical to elevating the impact of case study research in software engineering, particularly in its utility for practitioners and its theoretical contributions to the field.

\section{Observations and Takeaways}
\label{sec:observations}
One motivation for this work is a bittersweet experience when reading and reviewing case study research, particularly considering the industry application of the findings and the theoretical contributions to the field. 
While case studies inherently possess the potential to impact industry practices significantly, lacking crucial information that could make them more compelling and valuable for broader audiences often undermines this potential.
This gap between potential and realization has led us to reflect on some of the aspects presented in this paper. To better organize our thoughts and substantiate our observations, we have analyzed a sample of recently published case studies.
Specifically, we focused on papers published in 2022 and 2023 up to May of that year. The conferences encompass various topics within software engineering, such as Maintainability, AI Engineering, Project Management, Requirements Engineering, and CI/CD.

We examined a range of papers from various conferences, including three papers from ICSE 2023, one from ICSE SEIP 2023, one from ICSE CAIN 2023, one from ICSE CHASE 2023, one from ICSE 2022, one from REFSQ 2023, one from ICST 2023, and one from XP 2023.
While this selection of ten papers provides insight into current case study reporting practices in software engineering, it is important to clarify that our sample is neither statistically representative nor exhaustive.
Our choice to focus on recent conference papers is driven by the belief that these are more likely to be read by industry practitioners, hence having a significant practical impact. 
The space constraints of conference papers also provide insights into the authors' reporting priorities, yet they pose challenges in presenting comprehensive studies.
We intentionally avoid citing or referencing specific studies by anonymizing the papers in our analysis. Our goal is not to critique individual papers but to identify and exemplify broader patterns in case study reporting. This approach allows us to focus on general trends that could benefit from improvements, particularly in communication and applicability to academic research and industry practice.

Table \ref{tab:analysis_advice} summarizes our sample's analysis aspects and corresponding advice. Subsequent subsections detail these aspects.

\begin{table*}[!htbp]
  \centering
  \caption{Elements of Analysis and Corresponding Advice for Case Study Reporting}
  \label{tab:analysis_advice}
  \begin{tabular}{|p{0.2\linewidth}|p{0.7\linewidth}|}
  \hline
  \textbf{Element of Analysis} & \textbf{Takeaway} \\
  \hline
  Case Study Guidelines & Adhere to specific software engineering research guidelines like Runeson and Höst~\cite{runeson2009guidelines} and Ralph et al.~\cite{ralph2021acm} to ensure clarity, structure, and focus in the phenomenon. \\
  \hline
  Classification and Labeling & Employ the checklist by Wohlin and Rainer~\cite{wohlin2022case} to accurately classify research as a case study, enhancing the clarity and impact of your contributions. \\
  \hline
  Reporting the Context & Enhance case study reporting by descriptively illustrating the context using figures and using suggestions to contextualize research like Dyb{\aa} et al.~\cite{dybaa2012works} and Petersen et al.~\cite{petersen2009context}. \\
  \hline
  Generalizable Findings & Package contributions and articulate the generalizable findings of the case study, ensuring clarity on their applicability to different contexts. \\
  \hline
  Threats to External Validity & In addressing threats to external validity, focus on the generalizability of findings, how they build on existing theory, and discuss their applicability to broader contexts. \\
  \hline
  \end{tabular}
  \end{table*}

\subsection{Case Study Guidelines and Classification}
\label{sec:classification}

We observed a notable difference in the structure and clarity of case studies based on whether the authors cited and followed specific research guidelines. The guidelines often referenced were general research references for case study research~\cite{yin2003designing, merriam2015qualitative} and those specific to software engineering ~\cite{runeson2009guidelines, ralph2021acm}. Papers adhering to these methodological guidelines were better structured and more comprehensible. Notably, three of the five papers that did not follow these guidelines failed to present research questions explicitly. The lack of research questions makes understanding the study's motivation, methods, and results harder. These three studies were classified as small-scale evaluations, where the focus seemed less on studying a phenomenon in its authentic context and more on evaluating specific tools or processes. Regardless of the classification, following established guidelines can enhance the clarity and impact of case study research and facilitate a better understanding among reviewers and broader audiences by ensuring that critical elements are not overlooked.

\textbf{Takeaway:} 
\textit{By following established guidelines like those proposed by Runeson and Höst~\cite{runeson2009guidelines} or Ralph et al.~\cite{ralph2021acm}, researchers can verify that their studies are well-structured, clear, and focused on exploring phenomena in real-world contexts.}

We used the checklist proposed by Wohlin and Rainer to assess if the papers in our sample were authentic case studies~\cite{wohlin2022case}. Half of the papers were clearly identifiable as case studies, while the remainder showcased the classification complexity within software engineering research. Three papers were better suited under the label of ``small-scale evaluations'', one we were unsure of (probably not a case study), and one was definitively not a case study.

Using the checklist confirms the challenge of distinguishing among research categories. Whether a study is a case study, small-scale evaluation, or action research largely depends on the paper's focus and the authors' intentions.
Researchers should select the most appropriate classification, a decision that extends beyond mere labeling. Accurate classification is a fundamental step in enhancing the clarity of research communication.
A well-classified study makes it easier to communicate findings to a broader audience and assists reviewers in understanding the core contributions of the paper. 
Misclassification can lead to misunderstandings and potentially limit the scope and applicability of the research.
Furthermore, including terms like ``case study'' in the title can further clarify the research approach, aiding in accurate categorization and reader comprehension.

\textbf{Takeaway:} 
\textit{Employ the checklist by Wohlin and Rainer as a tool to determine if a study qualifies as a case study. This is not merely about labeling but is instrumental in enhancing the clarity and impact of the research contributions, facilitating a better understanding among reviewers and wider audiences.}

\subsection{Context Reporting}
\label{sec:classification}
We observed varying levels of context description within case studies. The context in small-scale evaluations tended to be more precisely defined, with authors often specifying tool combinations, practices, and technologies used. However, these detailed descriptions often led to the impression that the contexts were quite specific, potentially limiting the generalizability of the findings. In contrast, some real case studies appeared to lack certain aspects of context, making their generalizability and applicability more challenging to determine. To enhance the comprehensibility and relevance of context in case studies, we considered including visual aids like figures to illustrate the context. Similarly, following the suggestions by Dyb{\aa} et al.~\cite{dybaa2012works} and Petersen et al.~\cite{petersen2009context} can be highly beneficial to contextualize case studies. We noted that case studies that included figures illustrating the context were more easily comprehended. 

\textbf{Takeaway:}
\textit{
Utilize visual aids to depict the context, enhancing comprehension and accessibility of the case study.
Follow suggestions like those proposed by Dyb{\aa} et al.~\cite{dybaa2012works} and Petersen et al.~\cite{petersen2009context} to describe the context in a structured manner, ensuring a thorough and meaningful description of the case context.}

\subsection{Generalizable Findings and Threats to External Validity}
We noted varying degrees of clarity in articulating generalizable findings. While some papers included these findings within the introduction as contributions, others mentioned them in the discussion or conclusion sections. 
The critical factor, however, was a clear statement from the authors about what knowledge gained from the case study could be applied to other contexts. 
This clarity was notably missing in most papers, with only one explicitly outlining generalizable findings in the introduction. 
Particularly in small-scale evaluations, our interpretation of generalizable findings felt too generic, lacking specificity to the case context. 
Explicitly packaging generalizable findings in case studies helps readers understand the applicability of these learnings to other contexts.
Using technological rules suggested by Runeson et al.~\cite{runeson2020design} can be helpful.

\textbf{Takeaway:}
\textit{Clearly state the knowledge gained from the case study, ensuring that the findings are explicitly articulated and their applicability to other contexts is clearly stated.}

While most papers in our sample discussed threats to external validity, the effectiveness of these discussions was diverse. We encountered different interpretations, where some papers saw external validity as the influence of external factors on the case study. In the case of future studies and context broadening, papers reported as external validity the need for future research to expand the context rather than addressing the constraints of the current study. One paper dedicated a subsection to discuss the applicability of the findings to other contexts. Some other papers acknowledged the specificity of the context but failed to address how the results might apply to other contexts. Finally, one paper made commendable efforts, such as discussing generalizability/transferability, supporting findings with literature, and acknowledging the constraints of conference paper formats in providing detailed context descriptions.

We can conclude that the discussion of external validity does not consistently aid in understanding generalizable findings. There are various interpretations of external validity, ranging from those claiming case studies are not generalizable to those discussing the applicability of findings to other contexts. While not universally applicable, we agree that case studies can be analytically generalizable~\cite{runeson2009guidelines}. Additionally, from the design science perspective~\cite{runeson2020design}, findings may apply to similar problems.

\textbf{Takeaway:}
\textit{Effectively address threats to external validity by revisiting guidelines, building on existing theory, and discussing findings' generalizability and partial applicability.}

\section{Conclusion}
Summing up our exploration of case study reporting in software engineering, we highlight the importance of adhering to established research guidelines, proper classification of studies, detailed context description, clear articulation of findings, and meticulous addressing of generalizability threats. These elements, as detailed in our reflections, are crucial for enhancing the clarity, relevance, and impact of case studies. Our findings and recommendations, concisely captured in Table \ref{tab:analysis_advice}., offer actionable insights towards better case study reporting. This paper advocates for a shift in research practices to ensure that software engineering case studies are academically rigorous and practically applicable, bridging the gap between theory and industry application.

\bibliographystyle{ACM-Reference-Format}
\bibliography{references}

\end{document}